\documentclass[lettersize,journal]{IEEEtran}
\usepackage{amsmath,amsfonts}
\usepackage{algorithmic}
\usepackage{algorithm}
\usepackage{array}
\usepackage[caption=false,font=footnotesize,labelfont=rm,textfont=rm]{subfig}
\usepackage{textcomp}
\usepackage{stfloats}
\usepackage{url}
\usepackage{verbatim}
\usepackage{graphicx}
\usepackage{cite}
\usepackage{color}
\usepackage{amssymb}
\usepackage{multirow}
\usepackage{flushend}
\usepackage{tabularx}
\usepackage{soul}
\usepackage{upgreek}
\usepackage{textcomp}
\usepackage{makecell}
\usepackage[bb=boondox]{mathalpha}
\usepackage{hyperref} 
\hypersetup{
  colorlinks=true,
  citecolor=blue,
  linkcolor=red,
  urlcolor=green}

\newcommand{\bh}{\mathbf{h}}

\newcommand{\bb}{\mathbf{b}}

\newcommand{\bY}{\mathbf{Y}}
\newcommand{\by}{\mathbf{y}}

\newcommand{\bs}{\mathbf{s}}
\newcommand{\bN}{\mathbf{N}}
\newcommand{\bn}{\mathbf{n}}

\newcommand{\bH}{\mathbf{H}}

\newcommand{\bp}{\mathbf{p}}

\newcommand{\bI}{\mathbf{I}}
\newcommand{\bW}{\mathbf{W}}

\newcommand{\bx}{\mathbf{x}}
\newcommand{\bX}{\mathbf{X}}
\newcommand{\bF}{\mathbf{F}}
\newcommand{\bff}{\mathbf{f}}

\newcommand{\bz}{\mathbf{z}}

\newcommand{\sfT}{\mathsf{T}}

\newcommand{\calI}{\mathcal{I}}
\newcommand{\calV}{\mathcal{V}}
\newcommand{\CN}{\mathcal{CN}}

\newcommand{\bbC}{\mathbb{C}}

\newcommand{\bbE}{\mathbb{E}}

\newcommand{\ctrans}{\mathsf{H}}

\newcommand{\diag}{\mathrm{diag}}

\newcommand{\noisevar}{\mathcal{\sigma}_{\mathsf{n}}^2}

\newcommand{\UL}{\mathrm{ul}}
\newcommand{\DL}{\mathrm{dl}}

\newcommand{\pil}{\mathsf{p}}

\newcommand{\Nfft}{N_{\mathsf{fft}}}
\newcommand{\Nsc}{N_{\mathsf{sc}}}

\newtheorem{remark}{Remark}

\begin{document}

\title{Pioneering Scalable Prototyping for Mid-Band XL-MIMO Systems: Design and Implementation}

\author{
Jiachen Tian,~\IEEEmembership{Graduate Student Member,~IEEE,}
Yu Han,~\IEEEmembership{Member,~IEEE,}
Zhengtao Jin,
Xi Yang,~\IEEEmembership{Member,~IEEE,}
Jie Yang,~\IEEEmembership{Member,~IEEE,}
Wankai Tang,~\IEEEmembership{Member,~IEEE,}
Xiao Li,~\IEEEmembership{Member,~IEEE,}
Wenjin Wang,~\IEEEmembership{Member,~IEEE,} and
Shi Jin,~\IEEEmembership{Fellow,~IEEE}

\thanks{J. Tian, Y. Han, Z. Jin, J. Yang, W. Tang, X Li, W. Wang and S. Jin are with the National Mobile Communication Research Laboratory, Southeast University, Nanjing 210096, China (email: \{tianjiachen; hanyu; jinzhengtao; tangwk; li\_xiao; wangwj and jinshi\}@seu.edu.cn). X. Yang is with the Shanghai Key Laboratory of Multidimensional Information Processing, East China Normal University, Shanghai 200241, China, and is also with the National Mobile Communications Research Laboratory, South
east University, Nanjing 210096, China (e-mail: xyang@cee.ecnu.edu.cn). J. Yang is with the Frontiers Science Center for Mobile Information Communication and Security and also with the MOE Key Laboratory of Measurement and Control of Complex Systems of Engineering, Southeast University, Nanjing 211189, China (e-mail: yangjie@seu.edu.cn).}
}

\maketitle

\begin{abstract}
The mid-band frequency range, combined with extra large-scale multiple-input multiple-output (XL-MIMO), is emerging as a key enabler for future communication systems. 
Thanks to the advent of new spectrum resources and degrees of freedom brought by the near-field propagation, the mid-band XL-MIMO system is expected to significantly enhance throughput and inherently support advanced functionalities such as integrated sensing and communication. 
Although theoretical studies have highlighted the benefits of mid-band XL-MIMO systems, the promised performance gains have yet to be validated in practical systems, posing a major challenge to the standardization.
In this paper, preliminaries including frame structure, channel modeling, and signal models are first discussed, followed by an analysis of key challenges in constructing a real-time prototype system.
Subsequently, the design and implementation of a real-time mid-band XL-MIMO prototype system are presented.
Benefiting from the novel architecture, the proposed prototype system supports metrics aligned with standardization, including a bandwidth of 200 MHz, up to 1024 antenna elements, and up to 256 transceiver chains.
Operating in time-division duplexing (TDD) mode, the prototype enables multiuser communication with support for up to 12 users, while retaining standard communication procedures.
Built on software-defined radio (SDR) platforms, the system is programmable and allows for flexible deployment of advanced algorithms.
Moreover, the modular architecture ensures high scalability, making the system adaptable to various configurations, including distributed deployments and decentralized signal processing.
Experimental results with the proposed prototype system demonstrate real-time digital sample processing at 1453.33 Gbps, a peak data throughput of 15.81 Gbps for 12 users, and a maximal spectral efficiency approaching 80 bit/s/Hz.

\end{abstract}

\begin{IEEEkeywords}
6G, mid-band, OTA test, prototype system, XL-MIMO.
\end{IEEEkeywords}

\section{Introduction}

\IEEEPARstart{W}{ireless} communication technologies are evolving towards the sixth generation (6G), and increasingly stringent requirements are imposed in both performance metrics and functional capabilities.
As anticipated in \cite{6GProcIEEE,6GYou,6G}, 6G is expected not only to outperform previous generations in key metrics such as peak data rate, connection density, and spectrum efficiency, but also to support a variety of emerging applications, including integrated sensing and communication (ISAC), massive communication, among others.
As one of the primary objectives, enhancing transmission throughput necessitates concentrated endeavors in both the spatial and frequency domains, in accordance with the fundamental principles of wireless communications.

From the perspective of the spatial domain, massive multiple-input multiple-output (MIMO) technologies have received theoretical and engineering reputations due to the extraordinary capacities of spectral efficiency enhancement and interference suppression \cite{massiveMIMO,MaMIMOCM}.
On the one hand, theoretical analyses have proved the superiority of massive MIMO in exploring the spatial resource \cite{Ngo,MaMIMOJSAC,EmilTIT}.
On the other hand, towards the practical implementation, giant efforts have been made to investigate key technologies such as channel estimation, transmission strategies, etc. \cite{JSDM, Xie, SPMaMIMO}, and further incorporate the massive MIMO technologies into the whole procedures of a wireless communication system through corresponding standardization \cite{MaMIMOstandard}.

Despite the commercial success of massive MIMO, increasing the number of antenna elements remains a primary direction in the evolution of multi-antenna technologies. 
On this foundation, extra large-scale MIMO (XL-MIMO), equipped with enlarged arrays and flexible deployment schemes, is regarded as a promising regime \cite{JYZhang,HQLu}. 
Interestingly, preliminary studies indicate that emerging channel characteristics, particularly near-field propagation and spatial non-stationarities, are poised to dominate, unlocking a range of unforeseen advantages such as substantial improvements in throughput.
The novel channel characteristics and practical deployment are summarized in \cite{HanIOTJ}.
Starting from the channel characteristics, the authors in \cite{LuICC} identified the performance gains bounded by the near-field propagation characteristics.
In \cite{LDMA}, the extra degree of freedom (DoF) in the distance domain is explored.
Except for the enhancement in throughput, novelities in transmission design are also inspired by novel channel characteristics, such as transmission design \cite{XYangTVT}, and signal detection \cite{Amiri} with reduced computational complexity by utilizing the non-stationarities.
Furthermore, the introduced distance information calls for functionalities localization \cite{NFISAC,TianWCL}, thereby novel applications such as ISAC are inherently supported by XL-MIMO systems.

While advancements in the spatial domain continue to be explored, the limited existing spectrum resources remain a fundamental bottleneck. To break through the constraints of current spectrum allocation, both academia and industry are increasingly shifting the focus toward more promising frequency ranges. The World Radiocommunication Conference 2023 (WRC-23) issued Resolution 245, which suggests that the International Mobile Telecommunications (IMT) could potentially adopt the upper 6 GHz (U6G) frequency band, i.e., 6425-7125 MHz, globally \cite{WRC-23}. 
In comparison with the Sub-6 GHz and the millimeter wave (mmW), the mid-band spectrum showcases remarkable benefits, i.e., offering significantly wider bandwidth than the Sub-6 GHz band and exhibiting lower propagation loss compared to the mmW band. 
In addition, the mid-band frequency range also enables the integration of more antennas within the same array dimensions, compared to Sub-6 GHz systems \cite{GMIMO}. 
Consequently, the combination of mid-band frequency band and XL-MIMO has emerged as a key trend for future wireless communication systems \cite{WFanCM,TianTCOM}, promising enhanced spectral efficiency and the capability to support a massive number of users.

Although the aforementioned insightful theoretical studies have promised the superiority of the mid-band XL-MIMO system, the anticipated performance enhancements have yet to be measured.
There are only a few measurement campaigns focusing on the mid-band channel measurement and modeling \cite{MiaoJSAC, JHZhangCM, U6Garxiv}. 
To the best of the author's knowledge, suitable prototype testbeds for mid-band XL-MIMO system have not been proposed.
Recalling the research on massive MIMO in the fifth-generation (5G) era, groundbreaking practical validations and prototype systems were proposed prior to the successful commercial deployment \cite{XGao,argos,argosv2, LuMaMi,Lund, XYangChinaCom,Rapro}.
Specifically, the concept of massive MIMO was validated by measurement in \cite{XGao}, where the singular value spread and capacity are evaluated.
As for real-time prototyping, the Argos V1 \cite{argos} was built with a 64-antenna base station (BS), serving 15 single-antenna users simultaneously with 0.625 MHz of bandwidth in the TDD mode.
As an updated version, the Argos V2 \cite{argosv2} supported an increased number of BS antennas to 96 and 32 data streams.
The LuMaMi developed by the Lund University is a 100-antenna software-defined radio (SDR)-based centralized massive MIMO testbed capable of serving 12 single-antenna users in the same time-frequency resource with a TDD long term evolution (LTE)-like frame structure over a bandwidth of 20 MHz \cite{LuMaMi}.
Based on the testbed, measurement results of the time-varying channel are presented and analyzed in \cite{Lund}.
A time division duplex (TDD)-based 128-antenna massive MIMO prototype system was proposed in \cite{XYangChinaCom}, supporting an enhanced bandwidth of 20 MHz and throughput of 6.5 GByte/s.
Whilst in \cite{Rapro}, a novel architecture combined with field programmable gate arrays (FPGAs) and general purpose processors is proposed, enhancing the flexibility of algorithm development and deployment.
Although the aforementioned testbeds offer prospective designs for massive MIMO systems, they are not applicable for the validation of mid-band XL-MIMO systems due to the following reasons:
\begin{itemize}
    \item The existing architectures fail to meet the requirements in terms of novel mid-band, wide bandwidth in the hundreds of MHz, massive antenna elements and transceiver chains, and real-time signal processing capabilities, as dedicated in Table \ref{tab:cmp_system}.
    \item The system configurations are fixed, and flexible numerologies towards the 5G new radio (NR) and standardization of next generation cannot be realized for diverse validation scenarios.
\end{itemize}

Therefore, to align with the blueprint from an industrial perspective \cite{samsung,huawei,qualcomm}, we aim to develop a novel prototype system for the validation and standardization of the mid-band XL-MIMO system.
In this paper, we first review the preliminaries of mid-band XL-MIMO and outline key challenges associated with practical implementation.
The proposed prototype system in the typical mid-band frequency range of 6.4 to 7.2 GHz, features an enhanced bandwidth of 200 MHz, an extra large-scale array with up to 1024 antenna elements, and up to 256 transceiver chains.
The contributions of the paper are summarized as follows.
\begin{itemize}
    \item {\it Comprehensive preliminaries on the mid-band XL-MIMO system}. Key preliminaries are first reviewed, including frame structure, channel modeling, uplink and downlink transmission signal models, underpinning the theoretical basis of the proposed prototype system.
    Subsequently, the requirements of constructing the prototype system are extracted based on the analysis of practical challenges.
    \item {\it Detailed design and implementation of a real-time mid-band XL-MIMO system}.
    The illustration of the proposed prototype system encompasses the architecture of the BS side, the user side and the synchronization mechanism.
    The BS side and user side are specified in terms of the RF front-end, digital IF module, baseband MIMO processor module, and data interfaces.
    Benefiting from the novel architecture, the proposed system supports up to 1024 antenna elements and 256 transceiver chains with up to 200 MHz bandwidth.
    Besides, leveraging a modular design, the proposed prototype system exhibits high scalability, supporting various dimensional configurations (i.e., any number of transceiver chains from 1 to 256) and flexible deployment (i.e., centralized or distributed).
    Furthermore, advanced algorithms and functionalities can also be flexibly deployed and validated on the proposed prototype system, owing to the utilization of programmable hardware platforms.
    \item {\it Extensive experimental results for validating the proposed mid-band XL-MIMO system}. Based on the received signal, channel characteristics of mid-band XL-MIMO such as spatial non-stationarities are measured and analyzed.
    Configured with the proposed architecture, real-time constellation data transmissions in the uplink and downlink have been successfully realized.
    The successful real-time multiuser transmission demonstrates the feasibility and effectiveness of the proposed mid-band XL-MIMO architecture, underpinning research on transmission strategies, advanced algorithms and validation of practical performance.
    Additionally, the hardware performances of FPGAs on the BS side and user side are evaluated.
\end{itemize}

The rest of this paper is organized as follows. The preliminaries on mid-band XL-MIMO system are described in Section \ref{sec:XLMIMO}.
The system design and implementation of the prototype system are presented in Section \ref{sec:system}.
Section \ref{sec:result} presents the corresponding experimental results.
The conclusions are provided in Section \ref{sec:conclusion}.

{\it Notations}--Vectors and matrices are denoted by bold lowercase and uppercase letters, respectively.
The calligraphic letter $\mathcal{S}$ denotes a set.
The superscripts $(\cdot)^{\top}$ and $(\cdot)^{\mathsf{H}}$ represent the transpose and conjugate transpose, respectively;
$\diag(\cdot)$ represents a diagonal matrix. 
A complex Gaussian distribution is written as $\mathcal{CN}(\boldsymbol{\mu},\boldsymbol{\Sigma})$.

\section{Preliminaries on Mid-Band XL-MIMO Systems}
\label{sec:XLMIMO}

In this section, we outline the key preliminaries of mid-band XL-MIMO systems, covering system configuration, frame structure, channel modeling, and signal models.
These theoretical underpinnings serve as the basis for designing and implementing a viable mid-band XL-MIMO prototype system.

\subsection{System Configuration}
In this paper, a mid-band XL-MIMO system working in TDD transmission mode is considered, and the single-cell multiuser communication is focused. 
The BS is equipped with an $N$-element extra large-scale array, whilst $K$ single-antenna users are simultaneously served at the same time-frequency resource.
The orthogonal frequency division multiplexing (OFDM) technology is utilized, and the number of valid subcarriers and subcarrier spacing are denoted as $\Nsc$ and $\Delta f$, respectively.

\begin{figure}[!t]
    \centering
    \includegraphics[width=0.75\linewidth]{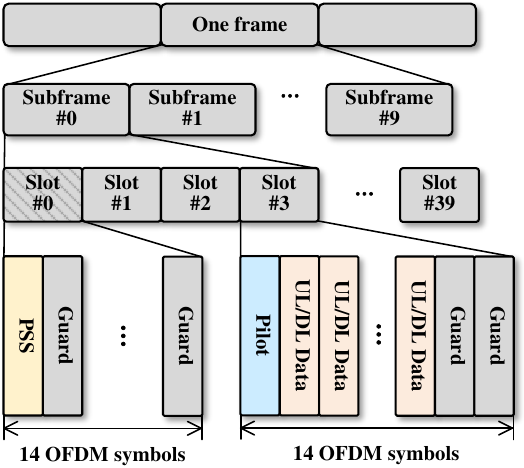}
    \caption{An example of the proposed frame structure for the prototype system under the subcarrier spacing $\Delta f = 60\  \mathrm{kHz}$, which is a simplified version of the 3GPP 5G NR frame structure.}
    \label{fig:frame}
\end{figure}

A simplified version of the 3rd Generation Partner Project (3GPP) 5G NR frame structure is adopted, where standard reference signals such as secondary synchronization signal (SSS) and demodulation reference signal (DMRS) in the synchronization signal block are reserved. 
Generally, a 10-ms radio frame is divided into 10 subframes.
As shown in Fig \ref{fig:frame}, we take $\Delta f=60$ kHz as an example for the illustration of frame structure, which is also aligned with the current prototype configuration (Refer to Table \ref{tab:system}).
Note that the frame structure in the proposed system is flexible due to high programmability, and thus various numerologies can be supported.
Each subframe is further divided into 4 slots, with each containing 14 OFDM symbols.
Except for slot 0, which is reserved for synchronization between the BS and the user via the primary synchronization signal (PSS) generated by a Zadoff-Chu (ZC) sequence, all other slots are utilized for pilot and data OFDM symbol transmission.
Within a slot, all data OFDM symbols can be dynamically allocated exclusively to either uplink or downlink transmission, thereby maximizing the data transmission rate and spectral efficiency. 
Considering the uplink and downlink switching in TDD mode, the transmission scheduling is performed at the slot level. Accordingly, 2 OFDM symbols are reserved as guard intervals, and the remaining 11 OFDM symbols can be allocated to either uplink or downlink data transmission except for a pilot symbol.
Given that the frame schedule in the proposed prototype system is software programmable, the frame structure can be flexibly reconfigured to accommodate different data transmission throughput and latency requirements of different scenarios.

\subsection{Channel Modeling}
Generally, wireless propagation can be modeled as a superposition of multiple propagation paths, which are further grouped into several clusters.
Different from massive MIMO systems, novel channel characteristics emerged in a mid-band XL-MIMO system, mainly including the near-field effects and spatial non-stationarities.
Therefore, traditional channel models require adaptation to accommodate the unique characteristics of mid-band XL-MIMO systems.
For a given user $k$, the channel vector on subcarrier $m$, $\bh _{k,m} \in \bbC ^{N \times 1}$, can be denoted as
\begin{equation}
    \bh _{k,m} = \sum _{s=1}^{S}\left( \sum _{\ell = 1} ^{L_s} \alpha _{k,s,\ell}\bb(\Omega _{k,s,\ell}) \right) \! \odot  \bp (\calV _{k,s}),
    \label{eq:channel}
\end{equation}
where $S$ and $L_s$ are the number of clusters and rays in cluster $s$, respectively, $\alpha_{k,s,\ell}$ represents the complex gain of the $\ell$-th ray in cluster $s$, resulted from path gain, initial phase, Doppler frequency and propagation delay, $\bb (\Omega _{k,s,\ell}) \in \bbC ^{N \times 1}$ is the near-field steering vector, with parameters $\Omega _{k,s,\ell} \triangleq \{r _{k, s,\ell},\ \theta _{k, s,\ell},\ \varphi _{k, s,\ell} \}$ reflecting distance, azimuth and downtilt of the $\ell$-th ray in cluster $s$ for user $k$, whose $n$-th entry is specified as
\begin{equation}
    [\bb(\Omega _{k,s,\ell})]_n =  \exp \left({\jmath {2 \pi (f_{c} + m \Delta f) D_{n}(\Omega _{k,s,\ell})}/{c} } \right),
\end{equation}
where $D_{n}(\Omega _{k,s,\ell})$ is the distance from polar coordinate $\Omega _{k,s,\ell}$ to the $n$-th array element, $f_c$ is the carrier frequency, and $c$ is the speed of light.
Regarding the spatial non-stationarities, $\bp(\calV _{k,s}) \in \bbC ^{N\times 1}$ is a weighted vector, whose elements are random variables (such as Bernoulli distribution) reflecting the non-stationarities in large-scale fading among different array elements at a cluster level, $\calV _{k,s}$ models the set of array elements with dominant channel power.

Observing the general channel model in \eqref{eq:channel}, parameters such as $r_{k,s,\ell}$ and $\calV _{k,s}$ are introduced due to emerging channel characteristics in mid-band XL-MIMO systems, which impact system performance and transmission design.
For instance, novel user scheduling and antenna selection schemes can be designed based on the spatial non-stationarities \cite{DFTXLMIMO,HQLu}, whilst a novel codebook is proposed utilizing near-field property \cite{LDMA}.
However, these studies focus on theoretical performance, lacking the practical validation.
Therefore, it is valuable to construct a mid-band XL-MIMO prototype system to validate system performances and novel algorithms.

\begin{figure}[!t]
    \centering
    \includegraphics[width=0.8\linewidth]{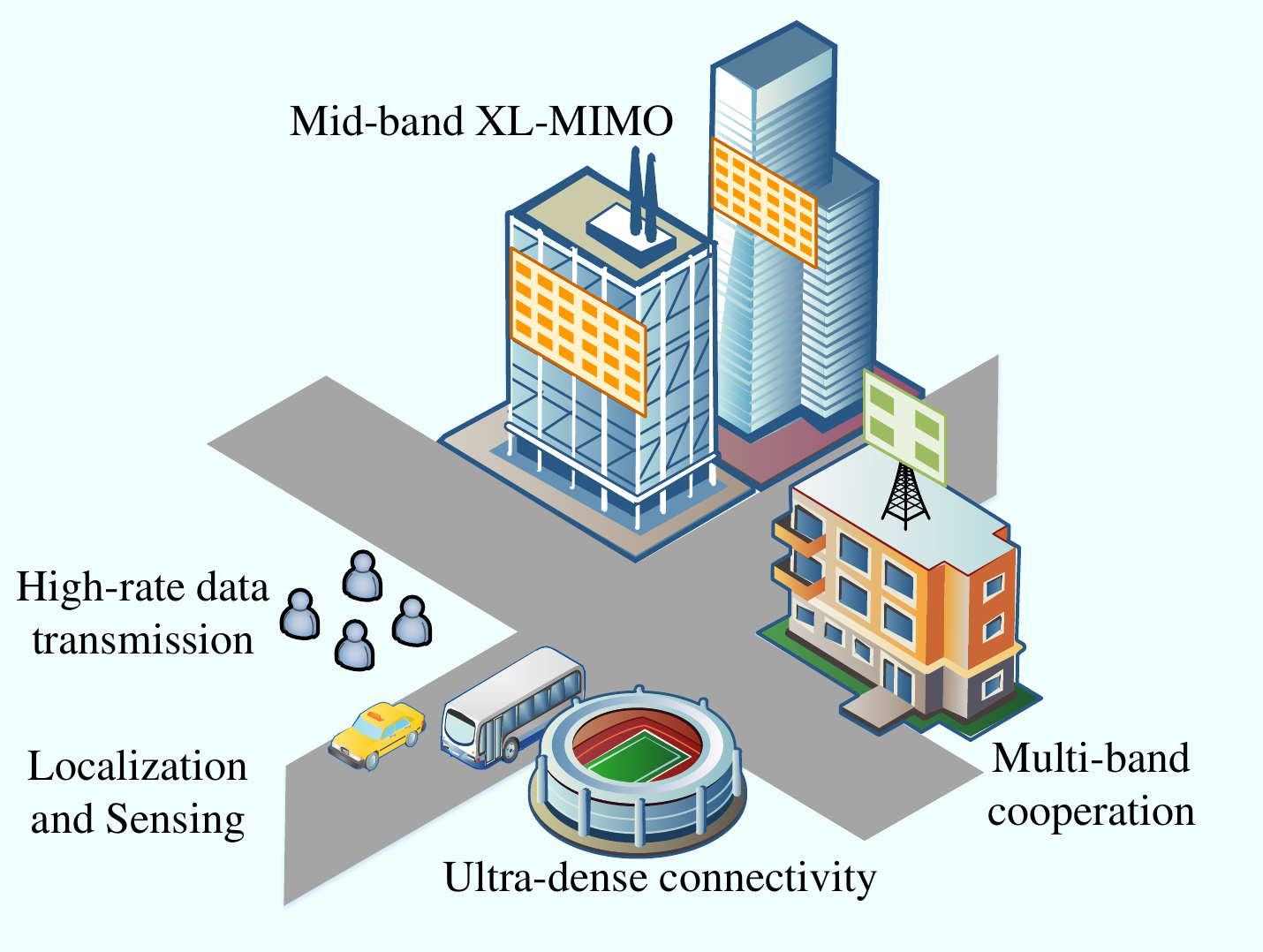}
    \caption{Mid-band XL-MIMO systems provide opportunities for high-rate data transmission, ultra-dense connectivity, localization and sensing, and multi-band cooperation.}
    \label{fig:XL-MIMO}
\end{figure}

\begin{table}[!t]
	\caption{Linear detection and precoding schemes \label{tab:PD}}
	\centering
	\begin{tabular}{m{0.4cm}<{\centering} c c c}
	\hline
	& MR & ZF & LMMSE \\
	\hline
        $\bW _{m}$ & ${\hat{\bH} _{m} ^{\UL}}{}^{\ctrans}$ & $({\hat{\bH} _{m} ^{\UL}}{}^{\ctrans}{\hat{\bH} _{m} ^{\UL}})^{-1}{\hat{\bH} _{m} ^{\UL}} {}^{\ctrans}$ & $({\hat{\bH} _{m} ^{\UL}}{}^{\ctrans}{\hat{\bH} _{m} ^{\UL}} + \noisevar\bI)^{-1}{\hat{\bH} _{m} ^{\UL}} {}^{\ctrans}$\\
        $\bF _{m}$ & ${\hat{\bH} _{m} ^{\UL}}{}^{*}$ & ${\hat{\bH} _{m} ^{\UL}}{}^{*} ({\hat{\bH} _{m} ^{\UL}}{}^{\ctrans}{\hat{\bH} _{m} ^{\UL}})^{-\sfT}$ & ${\hat{\bH} _{m} ^{\UL}}{}^{*} ({\hat{\bH} _{m} ^{\UL}}{}^{\ctrans}{\hat{\bH} _{m} ^{\UL}} + \noisevar \bI)^{-\sfT}$ \\
        \hline
	\end{tabular}
\end{table}

\subsection{Transmission Procedure}
\label{sec:signal}
In the proposed prototype system, the following transmission procedures are incorporated, and their underlying principles are introduced as follows.

\begin{figure}[!t]
    \centering
    \includegraphics[width=0.9\linewidth]{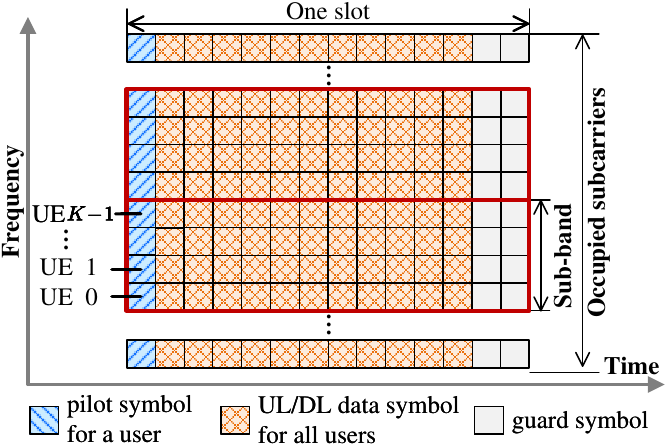}
    \caption{Time-frequency resource grids in TDD mode.}
    \label{fig:pilot}
\end{figure}

\subsubsection{Channel Estimation}
Before data transmission, channel estimation is carried out based on pilot OFDM symbols and we first take the uplink as an example.
As illustrated in Fig. \ref{fig:pilot}, frequency orthogonal pilots are adopted for different users, and $K$ neighboring subcarriers are successively allocated to $K$ users.
From the perspective of the frequency domain, every $K$ consecutive subcarriers are grouped into a sub-band, and $\Nsc/K$ sub-bands are obtained, denoted as $\calI = \{0,1,... ,\Nsc/K-1\}$.
User $k$ transmits pilots on the $k$-th subcarrier across all sub-bands, forming a comb type. That is, the channel can be estimated for user $k$ based on the pilots on subcarriers $iK+k$, $i\in \calI$. 
Specifically, the received pilot signal of the uplink pilot transmission on sub-band $i$ is denoted as
\begin{equation}
    \bY ^{\UL} _{\pil,i} = \bH _{i} ^{\UL} \bX ^{\UL} _{\pil,i} + \bN _{i},
    \label{eq:CEsignal}
\end{equation}
where $\bY^{\UL} _{\pil,i}\in \bbC ^{N\times K}$ is the received pilot signal at sub-band $i$, $\bH _{i} ^{\UL} \in \bbC ^{N \times K}$ is the uplink channel on sub-band $i$, $\bX ^{\UL} _{\pil,i} = \diag(x ^{\UL} _{\pil,i,0},\dots,x ^{\UL} _{\pil,i,K-1}) \in \bbC ^{K\times K}$ is the pilot signal matrix with the $k$-th column representing the pilot vector of user $k$,\footnote{In the prototype system, the pilots for different users are randomly generated in quadrature phase shift keying (QPSK) modulation.} and $\bN _{i}$ is the complex Gaussian noise matrix with elements satisfying independent identically distribution (i.i.d.) $\CN(0,\noisevar)$, where $\noisevar$ is the noise power.
Adopting the least squares (LS) channel estimation algorithm, the multiuser uplink channel on sub-band $i$ can be calculated as
\begin{equation}
    \hat{\bH} ^{\UL} _{i} = \bY _{\pil,i} ^{\UL} \bX _{\pil,i}^{\UL} {}^{\ctrans}.
\end{equation}
Afterwards, the channel estimates for users on the other $K-1$ subcarriers are derived based on the zero-order hold.\footnote{For more accurate channel estimation, frequency-domain interpolation can also be realized in the proposed prototype system at the cost of extra computational resources.}

For downlink channel estimation, the received downlink pilot signal at user $k$ on sub-band $i$ is
\begin{equation}
    y ^{\DL} _{\pil,i,k} = \bh _{i,k} ^{\DL} \mathbf{f}_{i,k} x ^{\DL} _{\pil,i,k} + n _{i,k},
\end{equation}
where $\bh ^{\DL} _{i,k} \in \bbC ^{1\times N}$, $\bff _{i,k} \in \bbC ^{N\times 1}$ and $x _{\pil,i,k} ^{\DL}$ are the downlink channel vector, the precoding vector, and the downlink pilot for user $k$ in sub-band $i$, respectively, and $n_{i,k}$ is Gaussian noise.
Multiplied by the $({x ^{\DL} _{\pil,i,k}})^*$, the downlink channel estimate is given by
\begin{equation}
    \hat{h} ^{\DL} _{i,k} = \bh _{i,k} ^{\DL} \mathbf{f}_{i,k} + n _{i,k} x ^{*} _{\pil,i,k}.
\end{equation}

\subsubsection{Uplink Data Transmission}
The uplink data transmission procedure follows the uplink channel estimation. 
Specifically, uplink data OFDM symbols are transmitted by $K$ users over the same time-frequency resource blocks.
Denoting the transmit signal for $K$ users on the $m$-th subcarrier as $\bs _m \in \bbC ^{K \times 1}$, the signal received at the BS over the $m$-th subcarrier is written as
\begin{equation}
    \by _m ^{\UL} = \sqrt{P ^{\UL}} \bH _m ^{\UL} \bs _m + \bz _m,
    \label{eq:yul}
\end{equation}
where $P ^{\UL}$ is the uplink transmit power, $\bH _{m} ^{\UL} \in \bbC ^{N\times K}$ is the uplink channel matrix on subcarrier $m$, $\bz _{m}\in \bbC ^{N\times 1}$ represents the complex noise with elements satisfying i.i.d. $\CN (0,\noisevar)$.
On the basis of channel estimates, the transmitted signal can be detected by
\begin{equation}
    \hat{\bs}_{m} = \bW _{m} \by _{m} ^{\UL},
\end{equation}
where $\bW _m \in \bbC ^{K\times N}$ is the detection matrix on subcarrier $m$.
Based on the channel estimates, $\bW _{m}$ can be designed according to the specific detection schemes, such as maximal ratio (MR), zero-forcing (ZF), and linear minimum mean square (LMMSE), are adopted, shown in Table \ref{tab:PD}.

\subsubsection{Downlink Data Transmission}
Reciprocal to the uplink transmission, the downlink transmission signal model can be expressed as
\begin{equation}
    \by _{m} ^{\DL} = \bH _{m} ^{\DL} \bF_{m} \bx _{m} + \bn _{m},
    \label{eq:ydl}
\end{equation}
where $\bH _{m} ^{\DL} \in \bbC ^{K \times N}$ is the downlink channel matrix on subcarrier $m$, $\bF _m \in \bbC ^{N\times K}$ is the downlink precoding matrix with specifications shown in Table \ref{tab:PD}, $\bx _{m} \in \bbC ^{K \times 1}$ is the information-bearing signals for $K$ users, satisfying the power constraint as $\bbE \{\bx_{m} ^{\ctrans} \bx_{m}\}=P ^{\DL}$, and $\bn _{m}$ is the additive complex noise with i.i.d. $\CN (0,\noisevar)$ elements.

At the user side, the received data is restored based on the channel estimate, i.e.,
\begin{equation}
    \hat{x} _{m,k} = ({\hat{h}^{\DL} _{m,k}})^{*} \bh _{m,k} ^{\DL}\left(\sum _{j=1} ^{K} \mathbf{f}_{m,j} \right) + ({\hat{h}^{\DL} _{m,k}})^{*} n _{m,k},
\end{equation}
where $\hat{h} ^{\DL} _{m,k}$ is the channel estimate on subcarrier $m$, which can be derived from $\hat{h} ^{\DL} _{i,k}$ via the zero-order hold.

\begin{remark}
    To align with the standard system, linear signal processing algorithms are adopted in the proposed system.
    Given the comprehensive signal models above, advanced algorithms can also be deployed on the proposed system.
    For instance, low-overhead localization can be applied to \eqref{eq:CEsignal}, whilst low-complexity signal detection algorithms can be deployed for \eqref{eq:yul} and \eqref{eq:ydl}.
\end{remark}

\subsection{Challenges and Requirements}
\label{sec:require}
As a promising and innovative paradigm for future BS architecture, the mid-band XL-MIMO system introduces a range of challenges stemming from its unique characteristics. 
Detailedly, according to industrial roadmaps, such systems are anticipated to support bandwidths in the range of 200 MHz to 400 MHz and antenna arrays comprising 256, 512, or even up to 1024 elements \cite{samsung,huawei,qualcomm}. Moreover, mid-band XL-MIMO is expected to integrate emerging technologies such as ISAC and ubiquitous connectivity, etc., shown in Fig. \ref{fig:XL-MIMO}. 
In light of these expectations, the proposed prototype system demands a {\it novel architecture} characterized by the following key features:
\begin{itemize}
    \item {\it Powerful computational capabilities:} The scaling-up of antenna elements and bandwidth introduces substantial computational challenges, notably in tasks like large-dimensional matrix inversion (related to $N$ and $K$) required by the LMMSE algorithm.
    The FPGAs exhibit significant advantages, including parallel computing capability, real-time responsiveness, low latency, and enhanced energy efficiency, which are suitable for constructing the prototype system.
    \item {\it High-speed data interfaces:} The expansions in bandwidth and data streams lead to a substantial increase in the volume of digital samples.
    As an example, considering the proposed system with the configurations provided in Table \ref{tab:system}, the total number of digital samples reaches
     \begin{equation}
        \frac{\Nsc\times28}{0.5\ \text{ms}}\times 16 \times 2(\mathrm{I/Q}) \times N = 1453.33\  \mathrm{Gbps},
        \label{eq:data}
    \end{equation}
    where $\Nsc=3168$, $N=256$ and the 16-bit analog-to-digital converters (ADCs) are assumed.
    To fulfill the real-time processing demands imposed by the large volume of digital samples, ensuring high-speed data interaction among various computational components is a critical aspect of the proposed system architecture.
    \item {\it Accurate synchronizatrion and triggering mechanism:} To support such a large system dimension, multiple devices are deployed on both the BS and user sides.
    Therefore, accurate synchronization and triggering mechanisms among a variety of devices are essential to enable simultaneous digital sampling and reliable signal processing. 
    \item {\it Exceptional scalability:} Owing to the enlarged array aperture, mid-band XL-MIMO systems support diverse deployment schemes, such as integration onto building surfaces. Consequently, the prototype system should adopt a modular design that is adaptable to not only distributed deployment in hardware, but decentralized signal processing as well.
    \item {\it Flexible and programmable development:} A primary objective of the prototype system is to validate the distinctive features of mid-band XL-MIMO, along with novel applications and functionalities.
    To this end, the system is required to provide high programmability and development efficiency, facilitating the implementation of advanced signal processing algorithms and enabling high-level functionalities such as localization and sensing.
\end{itemize}

\begin{table}[!t]\small
	\caption{System parameters of the prototype system \label{tab:system}}
	\centering
	\begin{tabular}{cc}
	\hline
	Parameter  & Value \\
	\hline
        \# of BS antennas $N$ & up to 1024\\
        \# of BS transceiver chains & up to 256\\
        \# of users $K$ & up to 12\\
        Baseband sample rate $F_s$ & 245.76 MS/s\\
        Carrier frequency & 6.4-7.2 GHz\\
        Subcarrier spacing $\Delta f$ & 60 kHz\\
        FFT size $\Nfft$ & 4096\\
        Data subcarriers $\Nsc$ & 3168 \\
        Occupied channel bandwidth & 190 MHz\\
        Min TDD switch duration & slot (250 $\upmu$s)\\
        \hline
	\end{tabular}
\end{table}

\section{Realization of the Proposed Prototype System}
\label{sec:system}
In this section, we detail the implementation of the proposed mid-band XL-MIMO prototype system, with key system parameters summarized in Table \ref{tab:system}. 
We begin by describing the selected hardware components, followed by the design from the perspectives of the BS side, user side and system synchronization. 
In addition, data transmission and signal processing across different hardware components are also discussed.

\begin{figure*}
    \centering
    \includegraphics[width=0.75\linewidth]{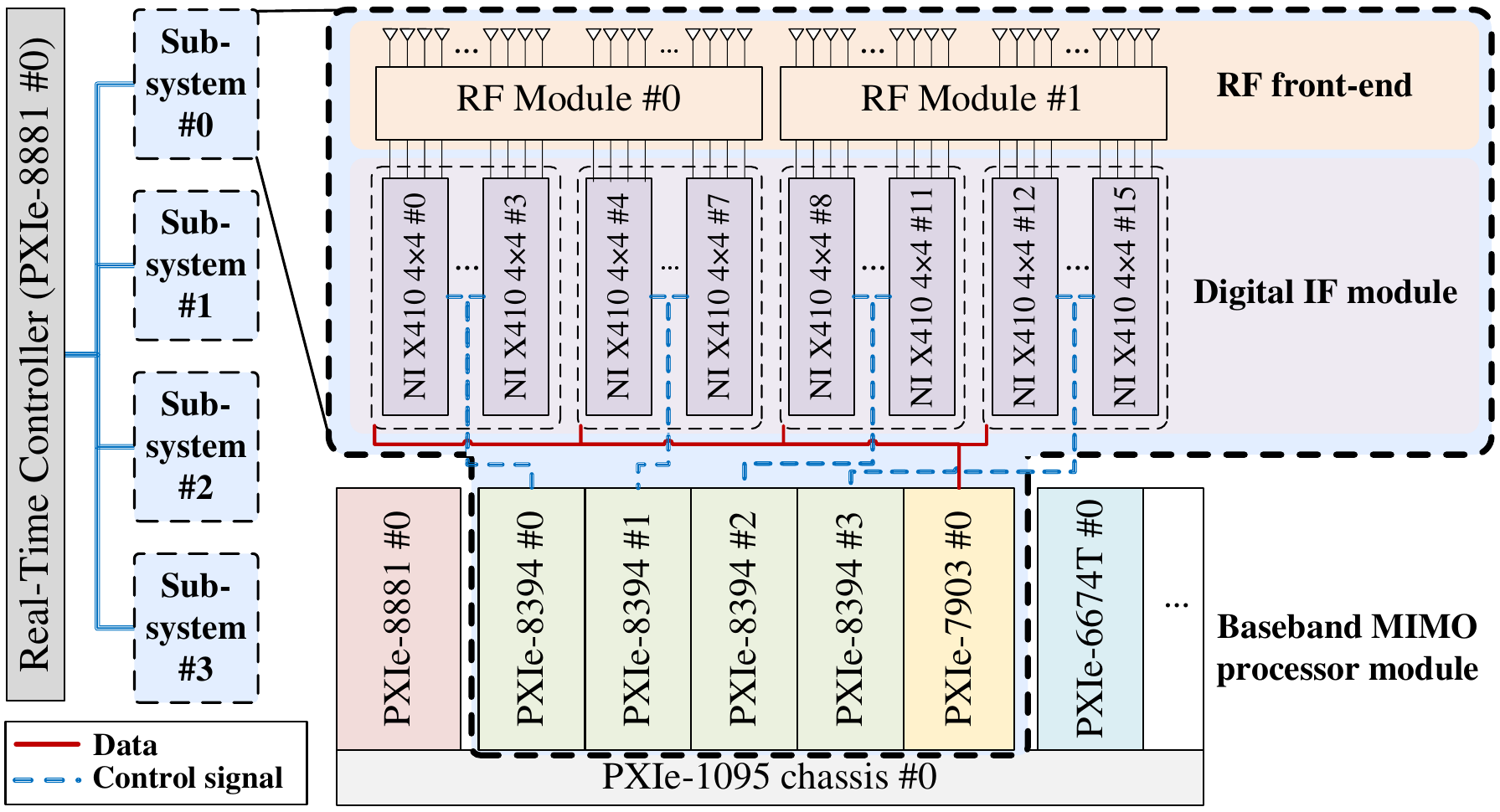}
    \caption{An example of the architecture of the BS side in the proposed prototype system with 256 transceiver chains.}
    \label{fig:BS}
\end{figure*}

\subsection{Selected Hardware Components}
To fulfill the requirements discussed in Section \ref{sec:require}, we construct the prototype system mainly by hardware components from National Instrument (NI), building on PCI Express Extensions for Instrumentation (PXIe) standard.
The detailed specifications are presented as follows.

\begin{itemize}
    \item {\it PXIe-1095 chassis:} 
    The PXIe-1095 chassis contains 18 slots, including 5 hybrid slots, 11 PXIe slots, and 1 PXIe system timing slot. 
    The PXIe-1095 chassis supports a maximal bandwidth of 24 GB/s and can be connected with PXIe-7903s through the PXIe-8881 controller. 
    \item {\it PXIe-8881:} A high-performance embedded controller based on Intel Xeon W-2225 processor with 4 cores and quad-channel 16 GB DDR4 memory. The maximal system bandwidth is 24 GB/s.
    \item {\it PXIe-7903:} Digital signal processor (DSP)-focused Xilinx XCVU11P FPGA co-processor, contains 9216 DSP slices and 12 mini-SAS HD interface, with each reaching a maximal data rate of 28.2 Gb/s.
    \item {\it Universal Software Radio Peripheral (USRP) X410:} SDR features a Xilinx ZU28DR radio frequency system-on-chip (RFSoC) with integrated FPGA and multi-core ARM processors.
    Four independent transceiver chains are included with a maximal bandwidth of 400 MHz and a center frequency ranging from 1 MHz to 7.2 GHz.
    The maximum transmitting power is 23 dBm.
    \item {\it PXIe-8394:} MXIe ×8 Gen 3 cabled PCIe interface suite, can be embedded into a PXIe chassis to converge data between the USRP X410 and PXIe-1095 chassis, with a real-time data transfer rate up to 7.9 GB/s.
    \item {\it PXIe-6674T:} Timing and trigger synchronization module with on-board highly stable 10 MHz oven-controlled crystal oscillator (OCXO). This module is used to generate the clock signal and enlarge the trigger signal, which can then be routed among multiple devices, such as PXI chassis and USRPs, to realize precise synchronization of timing and trigger signals across the whole system.
    \item {\it OctoClock-G CDA-2990:} The OctoClock-G CDA-2990 is used for high-accuracy timing and frequency reference distribution, which amplifies and distributes the internal or external 10MHz clock signal and trigger into 8 ways. 
\end{itemize}

\subsection{Base Station}
The architecture of the BS side in the proposed prototype system is shown in Fig. \ref{fig:BS}, showcasing a practical implementation with 256 transceiver chains.
It can be seen that the 256-transceiver chain implementation consists of four identical sub-systems connected to a Real-Time Controller, i.e., PXIe-8881.
The Real-Time Controller is further connected to a host computer through an Ethernet cable for program download and software debugging. 
From a functional perspective, the BS in the proposed prototype system can also be divided into the following components: the RF front-end, the digital intermediate frequency (IF) module, the baseband MIMO processor module, and data interfaces among them.
Specific illustration is presented as follows.

\begin{figure}[!t]
    \centering
    \includegraphics[width=0.75\linewidth]{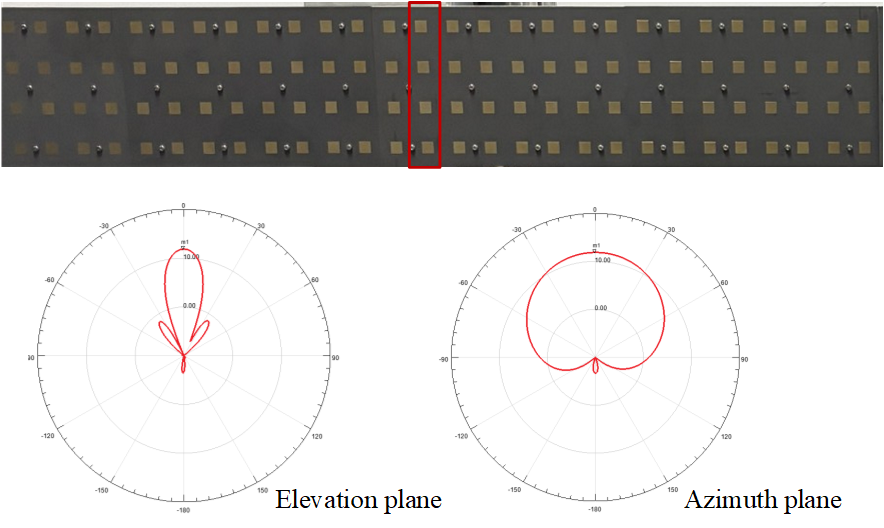}
    \caption{Part of the array consisting of 112 ($28\times 4$) elements and the radiation patterns at 6.75 GHz.}
    \label{fig:antenna}
\end{figure}

\begin{figure}[!t]
    \centering
    \includegraphics[width=0.85\linewidth]{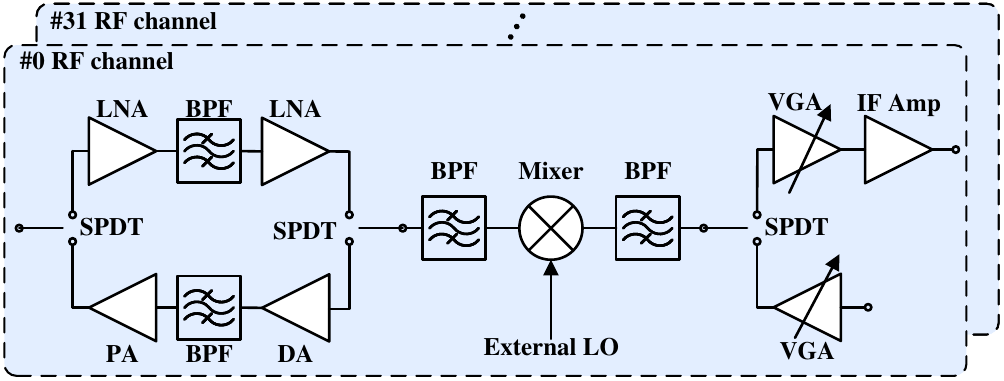}
    \caption{Internal architecture of the RF module.}
    \label{fig:RFmodule}
\end{figure}

\begin{figure*}
    \centering
    \includegraphics[width=0.8\linewidth]{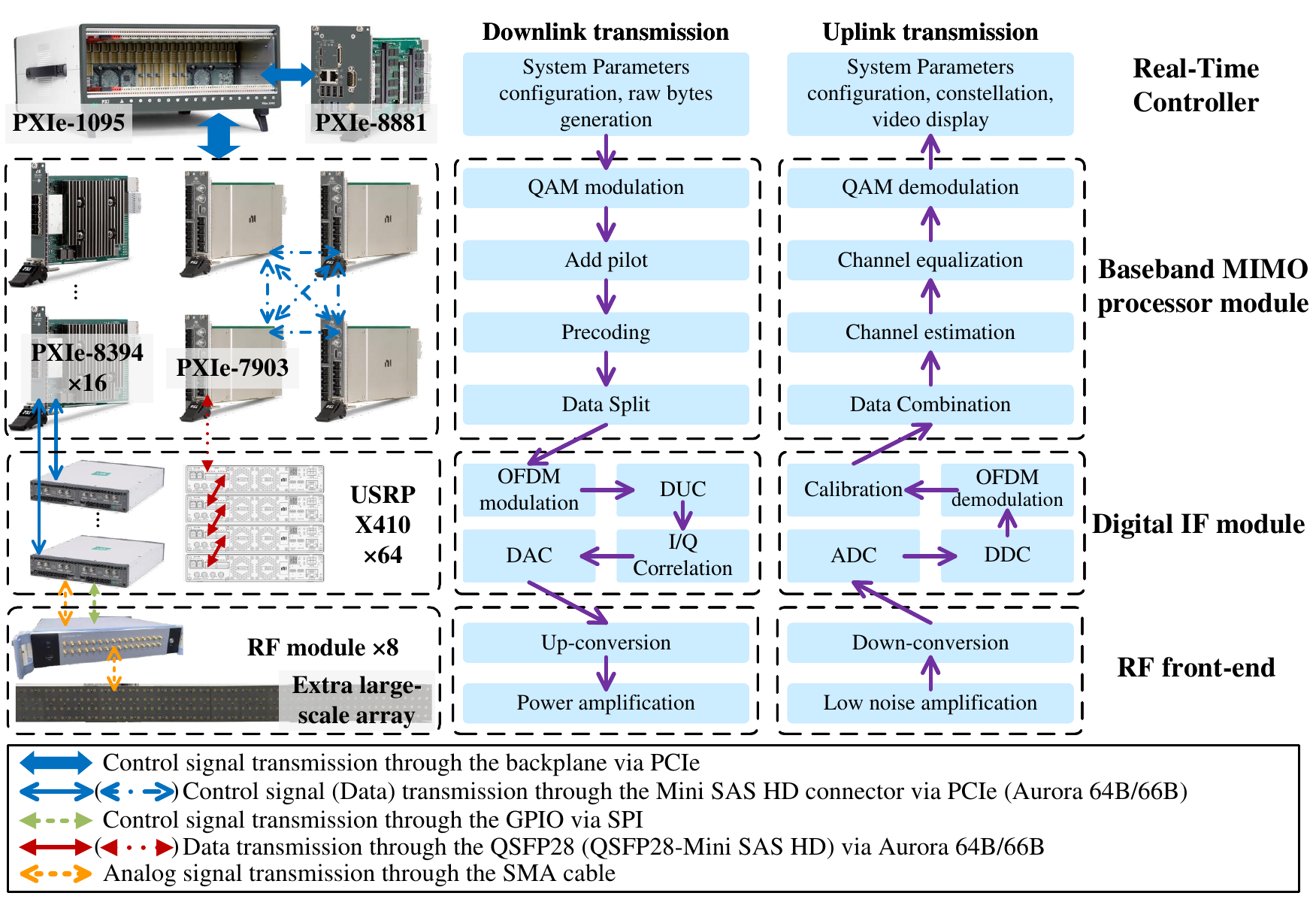}
    \caption{Data interfaces and signal processing procedures at the BS side in the proposed prototype system (uplink and downlink).}
    \label{fig:signalProces}
\end{figure*}

\subsubsection{RF Front-end}
The RF front-end at the BS side contains an extra large-scale array and corresponding RF modules.
The extra-large-scale array is configured as a uniform planar array (UPA) with 1024 elements, arranged with 256 elements horizontally and 4 elements vertically, with an inter-element spacing of $\lambda/2$ at 6.8 GHz.
Each array element is a microstrip antenna with linear polarization, and four elements in the vertical direction are directly combined to generate a narrow beam.
The resulting radiation patterns (of 4 elements in the vertical direction) in both the elevation and azimuth planes are shown in Fig. \ref{fig:antenna}, exhibiting 3-dB beamwidths of $20.76^{\circ}$ and $87.85 ^{\circ}$, respectively. 
As a result, the extra-large-scale array is capable of supporting up to 256 transceiver chains.
Note that the array adopts a modular design, with the basic unit being a subarray consisting of 32 elements arranged in an 8×4 configuration, facilitating both flexible reconfiguration in array shape and distributed deployment.

One RF module comprises 32 transceiver channels, each mainly integrating a mixer, power amplifiers (PAs), low-noise amplifiers (LNAs), bandpass filters (BPFs), variable gain amplifiers (VGAs), as shown in Fig. \ref{fig:RFmodule}.
In addition, single-pole double-throw (SPDT) switches for uplink/downlink switching are also adopted, enabling a switch speed $< 1 \upmu \mathrm{s}$ for the TDD mode.
The RF module supports programmable automatic gain control and automatic power control via VGAs. 
The supported radio frequency (RF) ranges from 6.4 to 7.2 GHz based on the input external local oscillator (LO), with an acceptable intermediate frequency (IF) range of 1.6 to 2.0 GHz. The RF module also accommodates a signal bandwidth of up to 400 MHz. 
In transmit mode, the output 1 dB compression point is 25 dBm, while in receive mode, the noise figure is below 3 dB.
The maximal transceiver channel gains under the transmit and receive modes are 45 dB and 55 dB, respectively.
To support an implementation with 256 transceiver chains, 8 RF modules are employed, whilst the array and the RF modules are connected through the SMA cables.

During uplink transmission, the received signal first passes through an LNA, followed by mixing with the LO for down-conversion.
Afterwards, the resulting analog IF signal is obtained after filtering and amplification by the IF amplifier.
Whilst in downlink transmission, the analog signal in each transceiver chain is first mixed with the LO for up-conversion, then passed through a filter and amplified by a power amplifier before being transmitted.

\subsubsection{Digital IF module}
To run up the total 256 transceiver chains, the digital IF module consists of 64 USRP X410 devices, each integrating 4 transceiver chains.
As shown in Fig. \ref{fig:signalProces}, for data interaction, every 4 X410 units are interconnected in a daisy-chain topology via Quad Small Form-factor Pluggable 28G (QSFP28) links, whilst 4 USRP X410 devices in a group are all connected to a PXIe-8394 for the control signal interaction with the controller through QSFP28 to Mini SAS HD connectors.

The digital IF module mainly carries out the conversion between the analog IF signal and the digital IF signal, and the OFDM modulation and demodulation, shown in Fig. \ref{fig:signalProces}.\footnote{Given the functionalities, the digital IF module can also be realized through the Xilinx RFSoC such as ZU47DR.
For instance, one Xilinx ZU47DR supports up to 8 transceiver chains, whilst providing a more efficient realization.
The realization based on the ZU47DR is left for our future consideration.}
During uplink transmission, the digital IF signal in a transceiver chain is first obtained via a 12-bit ADC, and then the high-rate samples are sent to the FPGA in USRP X410 for IQ imbalance correction, frequency shift correction, digital down conversion (DDC), and OFDM demodulation.
Afterwards, the obtained valid baseband data within a group of 4 USRP X410s is successively aggregated to one of them. 
For downlink transmission, the precoded data is distributed and transmitted to the USRP X410s for OFDM modulation.
After OFDM modulation, digital up conversion (DUC), frequency shift correction, and IQ imbalance correction are carried out in the FPGA of each URSP X410.
Afterward, the high-rate data bytes are conveyed to the 14-bit digital-to-analog converters (DACs) and the analog IF signals are obtained.

\subsubsection{Baseband MIMO processor module}
XL-MIMO-related baseband signal processing algorithms are implemented on the baseband MIMO processor module, which primarily consists of FPGAs.
As illustrated in Fig. \ref{fig:signalProces}, a total of 4 PXIe-7903 FPGA coprocessors are deployed to handle a digital signal with a 200 MHz bandwidth across 256 transceiver channels.
To deal with considerable digital samples, all PXIe-7903 coprocessors are interconnected via Mini SAS HD interfaces.
Due to the limited computational resources of each PXIe-7903, such interconnections between the PXIe-7903s enable data partitioning and aggregation flexibly.
In the implementation with 256 transceiver chains, each PXIe-7903 is in charge of a signal with 50 MHz bandwidth among 256 transceiver chains.
Whilst each PXIe-7903 can also deal with a signal with 200 MHz bandwidth among 64 transceiver chains, enabling distributed deployment and decentralized signal processing.

The uplink and downlink baseband signal processing procedures are given in Fig. \ref{fig:signalProces}.
For uplink transmission, the aligned baseband data are distributed to 4 FPGA co-processors, for LS channel estimation and QR decomposition-based LMMSE detection.
Before channel estimation and signal detection, data from every 16 USRP X410s is first exchanged among 4 FPGA coprocessors.
After such recombination, channel estimation and detection are carried out on the signal with 50 MHz among 256 transceiver chains, for each FPGA coprocessor.
Finally, the demodulated data are conveyed to the embedded controller by FPGA co-processors for display and further analysis.
Whilst in downlink transmission,  raw data bytes generated by the embedded controller for multiple users are first transferred to 4 FPGA co-processors. 
The precoding matrices used for downlink beamforming are pre-obtained during uplink channel sounding and are stored in the memories of the 4 FPGA coprocessors. Each FPGA co-processor is responsible for precoding the downlink data of 50 MHz bandwidth across 256 transceiver chains, using precoding matrices stored on board.
The precoded data based on the LMMSE algorithm in each PXIe-7903, is further exchanged with other PXIe-7903s.
After this interaction, the downlink data of 200 MHz bandwidth across 64 transceiver chains is obtained at each FPGA coprocessor and further conveyed to 16 USRP X410s.
It can be observed that the data interaction mechanism facilitates both distributed deployment and distributed signal processing. 
Moreover, identical algorithms can be deployed across all FPGA coprocessors, thereby endowing the system with high scalability.

\subsubsection{Data interfaces}
The internal data interfaces in the data IF modules and baseband MIMO processor modules have been illustrated in previous contents, and we mainly focus on the {\it data interfaces between different modules}.

For the data interface between the baseband MIMO processor module and the digital IF module, as calculated in \eqref{eq:data}, the data interaction remains a significant burden due to the increased bandwidth and transceiver chains.
To tackle this issue, the data interface between the baseband module and the digital IF module is realized as follows.
In the uplink transmission, the digital data from every 4 USRP X410 devices is aggregated and transmitted through a single QSFP28 to Mini SAS HD interface. This data stream is encapsulated using the Aurora 64B/66B protocol and delivered to the PXIe-7903 module. 
According to \eqref{eq:data}, the interaction throughput of every 16 USRP X410s is 90.83 Gb/s.
Therefore, the Multi-Gigabit Transceivers (MGTs) are configured to operate at 100\ \,Gb/s within each PXIe-7903, enabling high-throughput serial communication.
In the downlink transmission, the process is reversed: the high-speed data stream received by each PXIe-7903 is de-encapsulated and then distributed to 16 USRP X410 devices for further processing or playback.

The interaction of the control signal between the digital IF module and the RF front-end supports the TDD switch. 
The slot schedule instructions, generated by local frame timing, are preloaded at the USRP X410s.
The control signals are transmitted via the general-purpose input/output (GPIO) interface, where voltage levels are encoded according to a customized serial peripheral interface (SPI) protocol.
After completing the configuration triggered by the enable signal, the RF modules can be switched between uplink and downlink modes by corresponding voltage levels.
Combined with the accurate timing mechanism, seamless TDD switch among different transceiver chains can be guaranteed.

\subsection{User Equippment}
The user side contains antennas, RF modules, USRP X410s, PXIe-8394s, a PXIe-8881 and a PXIe-1095 chassis, which can be regarded as a simplified version of the BS.
Up to 12 single-antenna users are emulated using three USRP X410 devices.
For an arbitrary user, the antenna is connected to the RF module via an SMA cable, which is further linked to a transceiver channel of a USRP X410. 
The USRP X410 is interfaced with a PXIe-1095 chassis via a PXIe-8394 module. A PXIe-8881 controller is embedded within the PXIe-1095 chassis to manage the entire system.

The signal processing procedures are similar to the BS side, while data generation/recovery is implemented in the embedded controller or the computer, and the rest procedures are programmed in the FPGA contained in USRP X410s.

\subsection{Synchronization}
Timing and synchronization are critical issues for a system with multiple devices.
In the proposed prototype system, there exist two types of timing and synchronization issues.
\begin{itemize}
    \item Timing and synchronization between the BS and the users
    \item Timing and synchronization between devices at the BS or user side.
\end{itemize}

To tackle the former problem, the PSS generated by a 127-length ZC sequence is adopted.
Firstly, the BS transmits the PSS to the users, followed by the cross-correlation between the received PSS and local PSS at the user side.
Then, the index of peak among a 10-ms radio frame is obtained.
Given the unified frame structure shared at both the BS and the user side, the users transmit uplink data at the proper time.

\begin{figure*}[!t]
    \centering
    \includegraphics[width=0.85\linewidth]{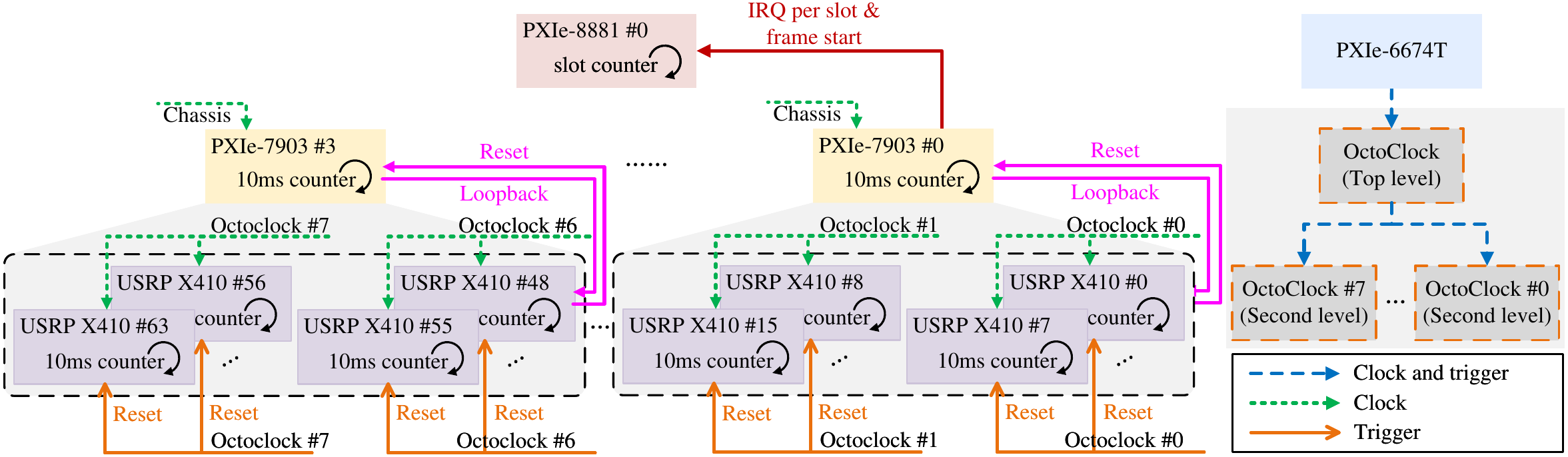}
    \caption{Clock and trigger distribution network.}
    \label{fig:clock}
\end{figure*}

As for the latter problem, a clock and trigger signal distribution network is established with the utilization of the OctoClock devices at the BS, shown in Fig. \ref{fig:clock}.
The OctoClock module is a signal amplifier and distribution module, using an external 10 MHz reference clock and an internal pulse per second (PPS) signal as a clock source and a trigger signal source, respectively. 
The principle of the clock and trigger signal distribution network can be summarized as follows: Firstly, the timing and synchronization module PXIe-6674T, which has an OCXO, generates a stable and precise 10 MHz reference clock locally.
The 10 MHz reference clock is subsequently forwarded to the top-level OctoClock module.
Combined with the generated local trigger signal, the top-level OctoClock module amplifies the received reference clock signal and local trigger signal and distributes them to 8 second-level OctoClock modules for further amplification and distribution. Finally, each second-level OctoClock module amplifies and distributes the reference clock signals and trigger signals to 8 USRP X410 devices. 
All USRP X410s are equipped with 10-ms counters, which are reset uniformly under the arrivals of PPS signals, and the frame synchronization among all USRP X410s can be realized.
For the PXIe-7903s, the clock signals are obtained from the chassis, whilst the trigger mechanism is as follows.
Backhaul interactions are established between USRP X410 units \#0, \#16, \#32, and \#48 and PXIe-7903 modules \#0, \#1, \#2, and \#3, respectively, where the 10-ms counters in PXIe-7903s are also uniformly reset by the arrival of trigger signal generated by these specific USRP X410s, thereby eliminating the need for dedicated clock and trigger signal connections.
Furthermore, a slot counter is equipped at the PXIe-8881, receiving the interrupt signal from the PXIe-7903 \#0.
Therefore, all 256 transceiver chains share the same reference clock signal and trigger signal, and all radio devices at the BS can start data collection and generation synchronously.
Benefiting from the mechanism, the dedicated links for clock and trigger signals from the digital IF module and baseband MIMO processor module are reduced, empowering higher scalability and flexibility.

\begin{table*}[!t]\small
\renewcommand{\arraystretch}{1.3}
	\caption{Comparisons between the proposed system with different prototype systems \label{tab:cmp_system}}
	\centering
	\begin{tabular}{m{2cm}<{\centering} m{1.5cm}<{\centering} m{1.6cm}<{\centering} m{2cm}<{\centering} m{1.3cm}<{\centering} m{1.2cm}<{\centering} m{1.5cm}<{\centering} m{1.5cm}<{\centering} m{1.5cm}<{\centering} }
	\hline
	 Systems  & \# of BS antennas & \# of transceiver chains & Carrier frequency & Bandwidth & FFT size & Modulation scheme & Supported standard & Scalability\\
     \hline
     Argos V2 \cite{argosv2} & up to 96 & up to 32 & 2.5-5 GHz & 625 kHz & N/A & N/A & N/A & conditional \\
     LuMaMi \cite{LuMaMi} & 100 & 100 & 1.2-6 GHz & 20 MHz & 2048 & up to 256-QAM & LTE & conditional\\
     System in \cite{XYangChinaCom} & 128 & 128 & 1.2-6 GHz& 20 MHz & 2048 & up to 256-QAM & LTE & conditional\\ 
     Rapro \cite{Rapro} & up to 128 & up to 128 & 1.2-6 GHz& 20 MHz & 2048 & up to 256-QAM & LTE & conditional \\
     Proposed & up to 1024 & up to 256 & 1.2-7.2 GHz &50/100/200 MHz & 1024/2048 /4096 & up to 256-QAM &LTE, NR& high\\ 
        \hline
	\end{tabular}
\end{table*}

\subsection{Discussions}
The proposed prototype system possesses the following advantages:
\begin{itemize}
    \item The proposed system holds a real-time mid-band XL-MIMO prototyping with 200 MHz bandwidth and 256 transceiver chains, due to powerful computational capabilities, high-capacity interfaces and accurate synchronization mechanisms.
    \item The prototype system demonstrates high scalability, supported by a modular design, the PXIe-based architecture, and a precise synchronization mechanism. 
    By flexibly combining different hardware modules, the system allows configurable dimensions, array topologies, and numbers of transceiver chains, thereby facilitating distributed deployment and decentralized signal processing.
    \item The proposed prototype system adopts a fusion computational architecture composed of SDR and FPGA.
    This design not only enables high-throughput data processing through the parallel computing capabilities of FPGAs, but also supports the deployment of advanced functionalities such as sensing and localization via the flexible programmability of SDRs.
    \item Comparisons with state-of-the-art massive MIMO prototype systems in the 5G era are listed in Table \ref{tab:cmp_system}.
    The proposed prototype system outperforms existing systems in key metrics such as bandwidth, number of array elements, and transceiver chains, closely aligning with the envisioned blueprint of mid-band XL-MIMO systems.
    Equipped with standard transmission procedure, the proposed system integrates the LTE and 5G NR alike, serving as a versatile testbed for validating the potential of mid-band XL-MIMO.
\end{itemize}

\section{Experimental Results}
\label{sec:result}

In this section, experimental results are presented and discussed.
Based on the experimental setup, the mid-band XL-MIMO channel measurement, multiple data stream uplink transmission, and multiuser beamforming data transmission in the downlink are demonstrated, validating the potential of mid-band XL-MIMO systems.
Furthermore, the hardware performance is also evaluated.

\begin{figure*}[!t]
    \centering
    \includegraphics[width=0.85\linewidth]{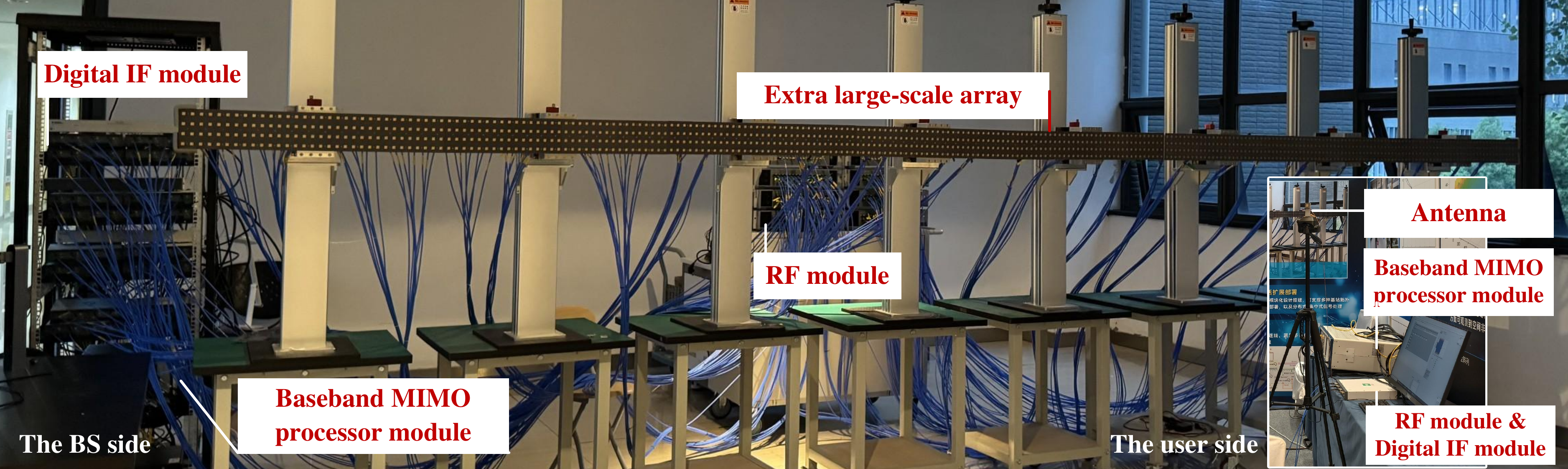}
    \caption{The implementation of the prototype system.}
    \label{fig:practicalSys}
\end{figure*}

\subsection{Experimental Setup}

\begin{figure}
    \centering
    \includegraphics[width=0.8\linewidth]{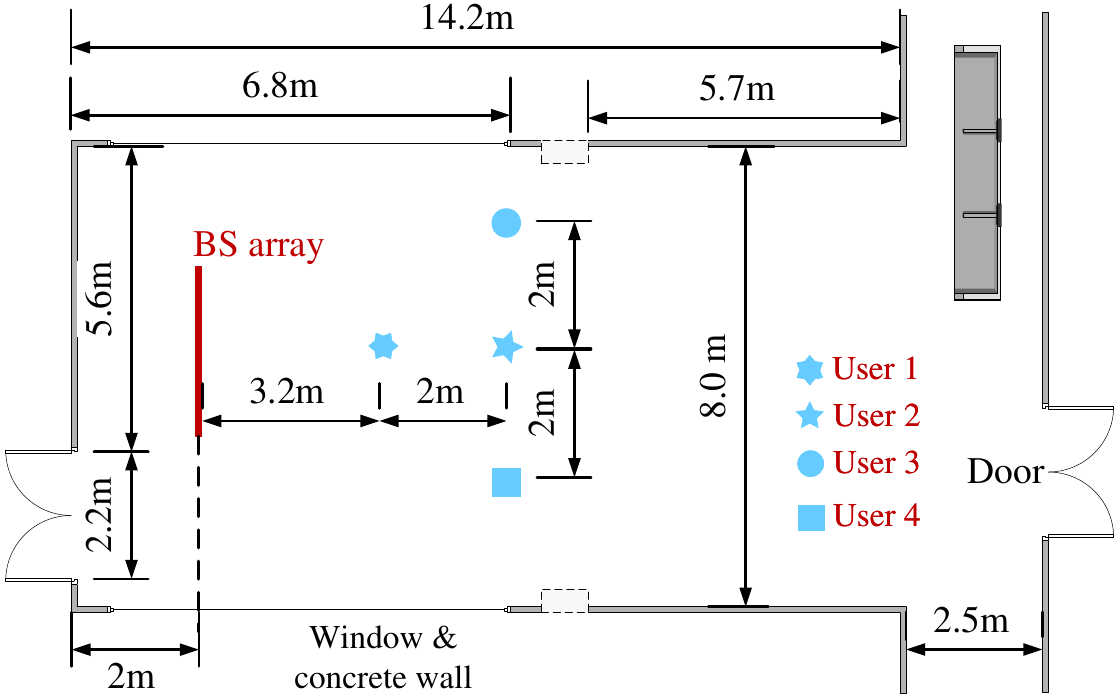}
    \caption{Layout of the indoor scenario for channel measurement. The Rayleigh distance is about 91 m.}
    \label{fig:scenario}
\end{figure}

During the measurement, we conduct an embodiment supporting 64 transceiver chains, as illustrated in Fig. \ref{fig:practicalSys}, whilst the number of transceiver chains can be extended up to 256 in future implementations.\footnote{As seen from Fig. \ref{fig:practicalSys}, the extra large-scale array and RF modules can support 256 RF transceiver chains. However, due to the limited number of devices in baseband MIMO processing module, the measurement results are presented based on an embodiment with 64 transceiver chains.}
The carrier frequency is configured as 6.8 GHz and other system configurations are the same as Table \ref{tab:system}.
The experiments are conducted in a typical indoor hall environment, and the deployment is presented in Fig. \ref{fig:scenario}.\footnote{As a step toward more realistic deployment, an outdoor scenario and experiments are under our future consideration.}
The indoor hall measures approximately 14.2 meters in length and 8 meters in width. Glass windows are installed on the upper halves of the two longer sides, while the shorter sides are constructed with concrete walls. The floor is covered with smooth ceramic tiles.
The array with 256 antenna elements arranged in a $64\times 4$ form is fixed in the center of the hall.
Whilst during the channel measurement, 4 single-antenna users are designedly positioned at different directions and distances relative to the array to verify the obtained DoFs in both he angular and distance domains brought by the mid-band XL-MIMO.
Specifically, user 1 and user 2 are aligned in the normal direction of the BS array, where user 1 is closer to the BS, while user 3, user 2, and user 4 are arranged at equal intervals in the direction parallel to the BS array.
For data transmission experiment, 8 single-antenna users are randomly distributed in front of the BS array.
Since the size of the array in the horizontal is $D = 63\lambda/2$, where $\lambda$ is the wavelength at 6.8 GHz, the Rayleigh distance is 
\begin{equation}
    d_{\mathsf{Rayleigh}} = 2D^2/\lambda \approx 87.5\ \mathrm{m},
\end{equation}
thereby all users are located in the near field.

\subsection{Channel Measurement}
\begin{figure}[!t]
    \centering
    \includegraphics[scale=0.5]{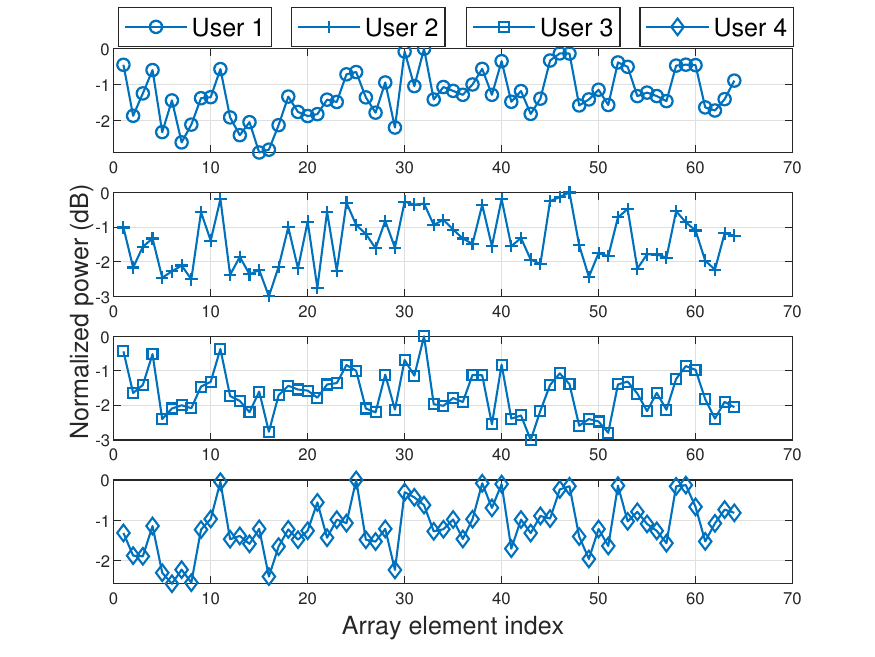}
    \caption{Normalized channel power versus array element index.}
    \label{fig:power}
\end{figure}

Based on the received uplink pilot signals, the BS estimates each user’s channel matrices by the LS channel estimation, configured with all-one pilots. 
Afterward, the measured channel matrices are further processed and analyzed, where the channel power and singular value spread are evaluated.
As shown in Fig. \ref{fig:power}, the normalized channel power distribution patterns of 4 users among 64 successive antennas are presented, where, for each user, the received power values of 64 antennas are normalized to their respective maximum.
Unequal power distribution can be observed, reflecting the spatial non-stationarities in the mid-band XL-MIMO channel.
Generally, the antenna elements far from the transmitter suffer more severe propagation loss, however, multipath effects also influence the channel power distribution. 
The observations validate the spatial non-stationary characteristics \cite{SNS}, and efficient transmission strategies can be derived by focusing on the array region with power concentration.

\begin{figure}[!t]
    \centering
    \includegraphics[scale=0.55]{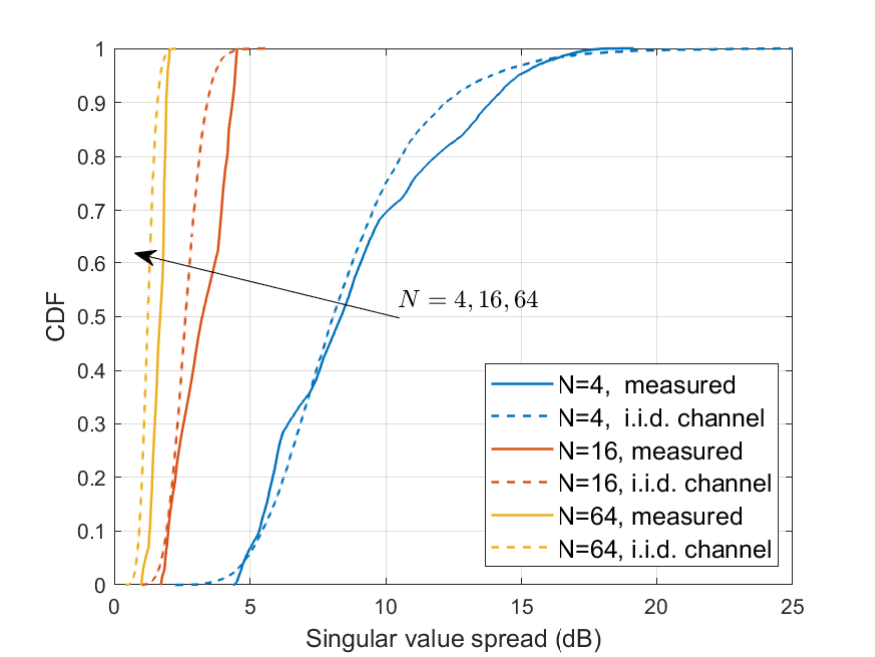}
    \caption{CDFs of singular value spread for 4, 16 and 64 antennas.}
    \label{fig:singular}
\end{figure}

To investigate the favourable propagation, the cumulative distribution functions (CDFs) of the singular value spread are presented in Fig. \ref{fig:singular}, with comparisons conducted for array configurations of 4, 16, and 64 elements.
The singular value spread is defined as the ratio of the maximum to the minimum singular value, obtained from the singular value decomposition (SVD) of the channel matrix formed by the channel vectors of 4 users.
Note that the channel vector of each user is normalized \cite{XGao}.
It can be observed that the favourable propagation is guaranteed by the DoF obtained in both the angular and distance domains, since the users are not only in different directions, but also in the same direction with different distances.
As the number of array elements increases, the singular value spread reduces towards 1, thereby creating a favorable propagation for multiuser communication.
The measured singular value spreads are close to those under the i.i.d. Gaussian channel, which further validates the capabilities of multiuser communication simultaneously.

\subsection{Uplink and Downlink Transmission}

\begin{figure}[!t]
    \centering
    \includegraphics[width=0.75\linewidth]{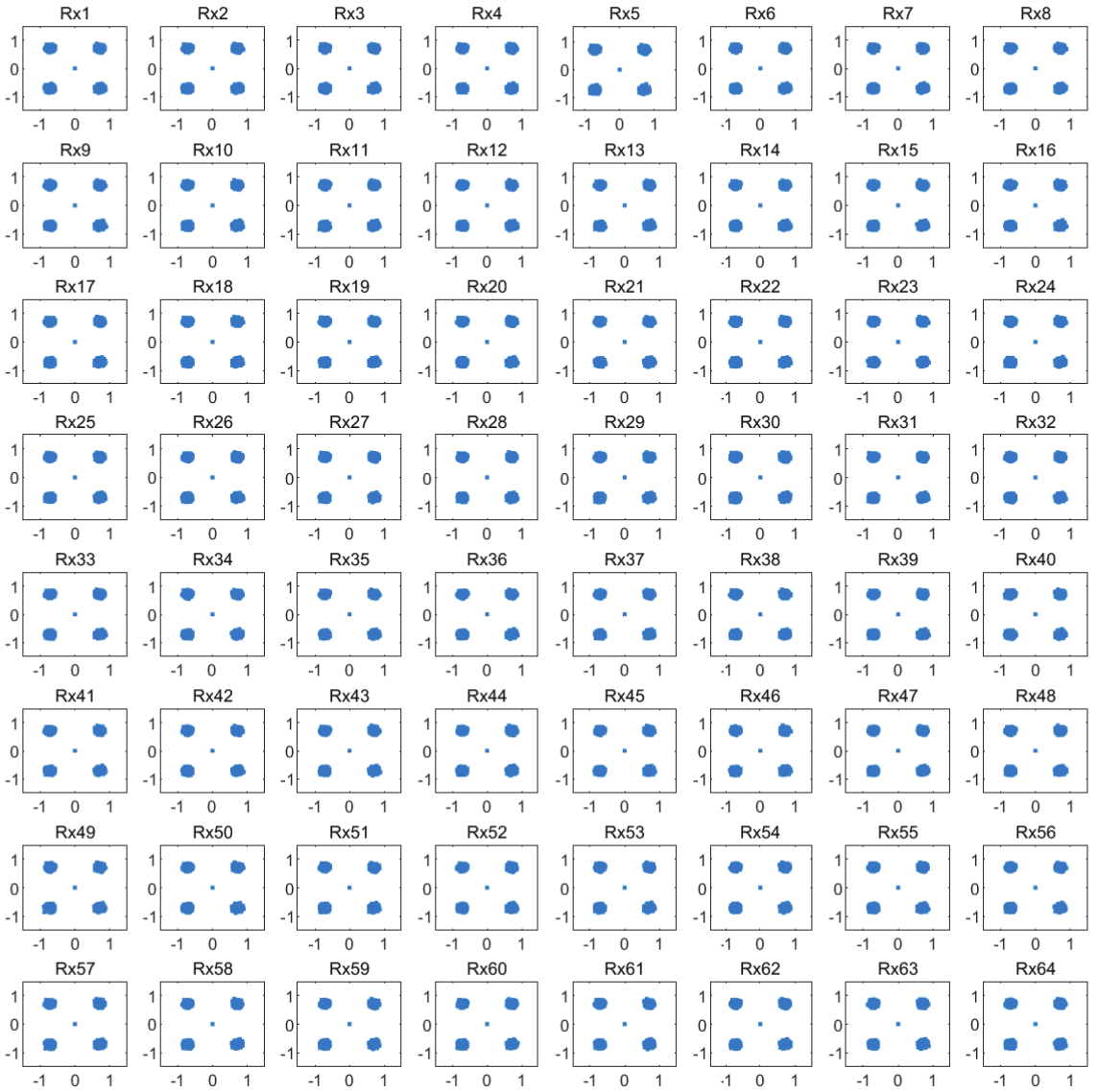}
    \caption{The constellation of detected symbols of 64 transceiver chains in uplink transmission.}
    \label{fig:constellation_64}
\end{figure}

\begin{figure}[!t]
	\centering
	\subfloat[Uplink constelation transmission]{\includegraphics[width=.7 \linewidth]{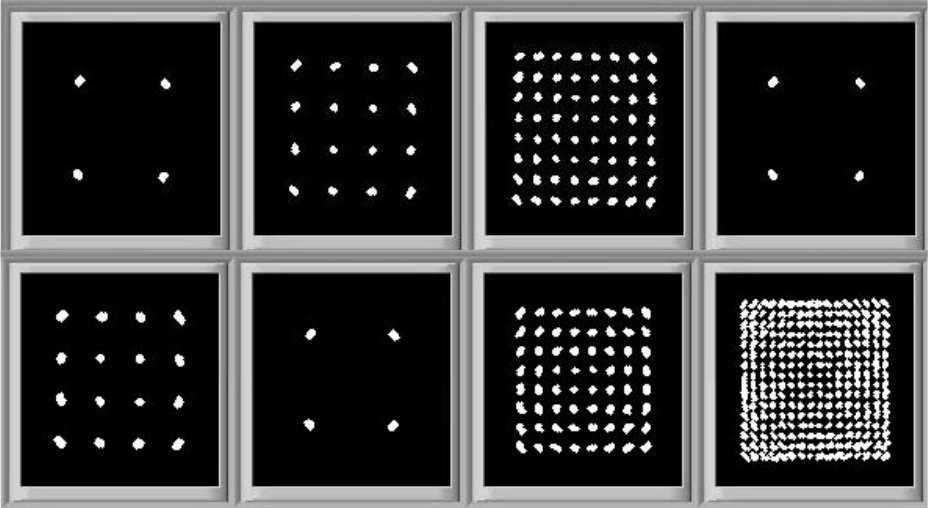 }\label{fig:constellation_1}}\\
	\subfloat[Downlink constellation transmission]{\includegraphics[width=.7 \linewidth]{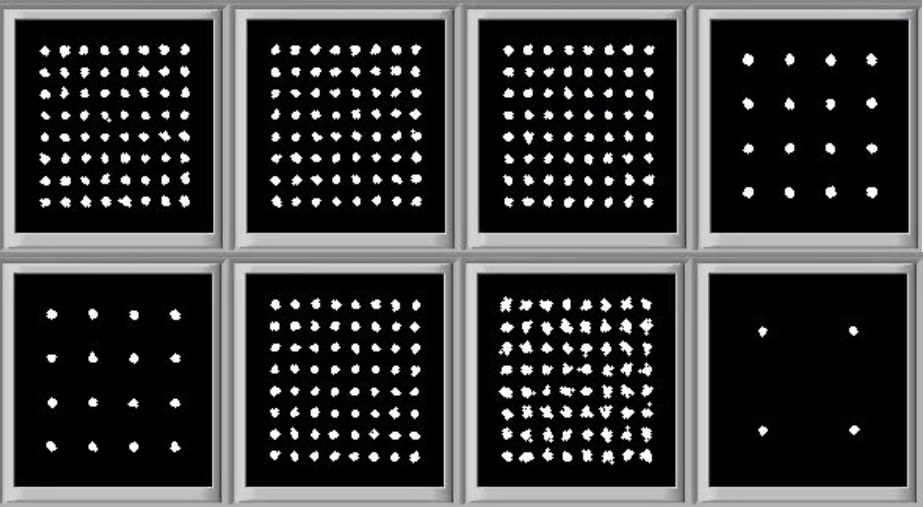} \label{fig:constellation_2}}\\
    \caption{Performance of constellation in uplink and downlink.}
    \label{fig:constellation}
\end{figure}

Equipped with the configured frame structure provided in Fig. \ref{fig:frame}, the uplink and downlink data transmission experiments are conducted.
Firstly, the constellations of 64 transceiver chains in uplink transmission are shown in Fig. \ref{fig:constellation_64}, where a single-antenna user transmits data to the BS with the QPSK modulation scheme. The 64 transceiver chains at BS successfully recover the data streams, which validates the performance of the 64 transceiver chains.
As for the real-time constellation transmission, the restored constellation for eight users in uplink is shown in Fig. \ref{fig:constellation} \subref{fig:constellation_1}.
If all OFDM symbols are allocated for data transmission, the theoretical peak throughput can be calculated as
\begin{equation}
\begin{aligned}
    \frac{\Nsc\times 26 (\mathrm{symbols}) \times 8 (\mathrm{256\mbox{-}QAM}) \times 8(\mathrm{users}) }{500\times 10 ^{-6}(\mathrm{half\ sub\mbox{-}frame}) } &\\
    = 10.56\  \mathrm{Gbps},&
\end{aligned}
\end{equation}
when the 256-QAM modulation are adopted for all users.
Whilst the peak throughput reaches
\begin{equation}
\begin{aligned}
    \frac{\Nsc\times 22(\mathrm{symbols}) \times 8(\mathrm{256\mbox{-}QAM}) \times 8(\mathrm{users}) }{500\times 10 ^{-6}(\mathrm{half\ sub\mbox{-}frame})} &\\
    = 8.92\  \mathrm{Gbps},&
\end{aligned}
\end{equation}
under consideration of the TDD switch.
The downlink constellation is illustrated in Fig. \ref{fig:constellation} \subref{fig:constellation_2}, the 256-QAM modulation scheme is used for five of the eight single-antenna users, and the other three users adopt 16-QAM and QPSK, which manifests an achieved peak throughput of 6.59 Gbps over 200 MHz bandwidth.
When serving 12 users, the maximum throughput can reach 15.81 Gbps, outperforming state-of-the-art prototype systems for massive MIMO systems \cite{XYangChinaCom}.
In addition, the maximum spectral efficiency of the proposed prototype system reaches
\begin{equation}
\begin{aligned}
    \frac{\Nsc \!\times \! 26 (\mathrm{symbols}) \! \times \! 8 (\mathrm{256\mbox{-}QAM}) \! \times \! 12(\mathrm{users}) }{500 \! \times \! 10 ^{-6} (\mathrm{half\ sub\mbox{-}frame}) \! \times \! 200 \! \times \! 10^6 (\mathrm{bandwidth})}& \\
    = 79.12\  \mathrm{bit/s/Hz}.&
\end{aligned}
\end{equation}
Note that the maximum spectral efficiency can still be improved by increasing the valid subcarriers from a 4096 FFT size. For example, we can use 3276 of the 4096 subcarriers to transmit data and thus achieve a maximum spectral efficiency of 81.82 bit/s/Hz.
Besides, benefiting from the near-field DoF and non-stationary properties, more users can be simultaneously supported based on novel precoding and detection algorithms, thereby further enhancing the spectral efficiency.

\begin{table*}[!t]\small
\renewcommand{\arraystretch}{1.3}
	\caption{Hardware utilization of FPGAs at the BS side and user side of the proposed mid-band XL-MIMO prototype system \label{tab:hardware}}
	\centering
	\begin{tabular}{m{2cm}<{\centering} m{2.4cm}<{\centering} m{2.2cm}<{\centering} m{2cm}<{\centering} m{2.4cm}<{\centering} m{1.5cm}<{\centering} m{2cm}<{\centering} }
	\hline
     & Slice Registers & Slice LUTs & Block RAMs & UltraRAM Blocks & DSP48s & Total Slices \\
     \hline
     \multirow{2}*{\makecell[c]{BS side \\(per PXIe-7903)}} & 599922/2592000 & 340743/1296000 & 591/2016 & 572/960 & 7720/9216 & 92535/162000 \\
     & 23.7\% & 26.3\% & 29.3\% & 59.6\% & 83.8\% & 57.1\% \\
     \hline
     \multirow{2}*{\makecell[c]{BS side\\(per USRP X410)}} & 338491/850560 & 189738/425280 & 432/1080 & 54/80 & 2734/4272 & 42909/53160 \\ 
     & 39.8\% & 44.6\% & 40.0\% & 67.5\% & 64.0\% & 80.7\% \\ 
     \hline
     \multirow{2}*{\makecell[c]{User side\\(per USRP X410)}} & 360029/850560 & 207938/425280 & 525/1080 & 6/80 & 3017/4272 & 47758/53160 \\ 
     & 42.3\% & 48.9\% & 48.6\% & 7.5\% & 70.6\% & 89.8\% \\ 
     \hline
	\end{tabular}
\end{table*}

\subsection{Hardware Performance Analysis}
The performances of hardware resource utilization for the implementation of the proposed prototype are given in Table \ref{tab:hardware}.
The FPGAs equipped at the baseband MIMO processor module, digital IF module of the BS side, and the user side are evaluated.
Note that the hardware resource utilization is identical across different PXIe-7903s and across different USRP X410s, which proves that the system can be reconfigured flexibly and achieve enhanced system robustness.
Given the burden task of performing LS channel estimation, LMMSE detection and LMMSE precoding, more than half of the hardware resources are used in the FPGAs.
For instance, the utilization ratios of the used DSP48E at the baseband MIMO processor module and the digital IF module side are 83.8\% and 64.0\%, respectively, mainly resulting from the pseudo-inverse computation of the $256\times 12$ matrix and the OFDM modulation/demodulation, respectively.
The considerable hardware resource usage severely hinders the introduction of error correction codes, such as turbo or polar codes, whose implementation necessitates extra FPGAs. Therefore, advanced detectors, which fully exploit massive MIMO system characteristics with less realization complexity, should be introduced. 
A robust channel estimator such as LMMSE channel estimator, also needs to be considered to acquire more accurate channel state information.
Due to the data routing, data alignment and data storage for debugging, a significant portion of the RAM resources is also consumed.

\section{Conclusion}
\label{sec:conclusion}
Motivated by the promising integration of mid-band spectrum and XL-MIMO technologies, the design and implementation of a mid-band XL-MIMO prototype system are presented in this paper.
The proposed prototype system supports a 200 MHz bandwidth, up to 1024 array elements, and up to 256 transceiver chains.
Operating in TDD mode, the prototype system enables programmable configurations for both uplink and downlink transmissions. 
Leveraging a modular design, the proposed system offers high scalability, enabling flexible adaptation to distributed deployment and decentralized signal processing, catering to the practical implementation of mid-band XL-MIMO systems.
In addition to linear signal processing algorithms, advanced algorithms supporting novel functionalities such as ISAC can be deployed on the proposed prototype, due to the utilization of an SDR-based framework.
The proposed prototype system validates the performance of mid-band XL-MIMO through extensive experimental results.
The spatial non-stationarities are observed, and the favourable propagation conditions are also verified.
Except for analyzing novel channel characteristics, the proposed prototype system realizes a real-time uplink and downlink transmission, with the maximum spectral efficiency of 79.12 bit/s/Hz.
Furthermore, the hardware utilization performance is also evaluated.

Conclusively, the proposed TDD-based 256-transceiver chain mid-band XL-MIMO prototype system provides a sufficient but scalable reference design for research on the mid-band XL-MIMO system. 
Future works will be directed toward introducing an LMMSE channel estimator and advanced detectors in the proposed system. 
Whilst advanced baseband processing techniques, especially utilizing channel characteristics, will be deployed and validated in our system.
Considering the hardware cost of the current version, efficient realizations based on the Xilinx RFSoC will also be considered in the near future.

\bibliographystyle{IEEEtran}
\bibliography{IEEEabrv,XLMIMOsystem}

\end{document}